\documentclass[aps,prl,reprint,groupedaddress]{revtex4-2}
\usepackage{graphicx}
\usepackage{siunitx}
\usepackage{color}

\usepackage{amsmath}
\usepackage{amssymb}
\usepackage{mathtools}
\usepackage[english]{babel}

\newcommand{\vect}[1]{\boldsymbol{#1}}

\begin{document}


\title{Multiphoton ionization with three-dimensional light fields}


\author{D. Köhnke}
\thanks{These authors contributed equally to this work.}
\author{H.-C. Ahlswede}
\thanks{These authors contributed equally to this work.}
\author{T. Bayer and M. Wollenhaupt}
\email[]{matthias.wollenhaupt@uol.de}
\affiliation{Carl von Ossietzky Universität Oldenburg, Institute of Physics, Carl-von-Ossietzky-Straße 9-11, D-26129 Oldenburg}

\date{\today}

\begin{abstract}
We report the first observation of free-electron angular momentum wave packets generated by atomic multiphoton ionization with bichromatic three-dimensional (3D) polarization-tailored ultrashort laser fields. These fields, created by the non-collinear superposition of two polarization-shaped pulses of different colors from a supercontinuum polarization pulse shaper, provide electric-field components along all spatial directions. The resulting photoelectron momentum distributions, recorded via velocity map imaging, demonstrate full 3D coherent control of electronic superposition states extending beyond the constraints of planar polarization fields by unlocking all dipole selection rules $\Delta m = 0,\pm1$. As an application, 3D pump–probe fields are used to image previously unobserved photoelectron wave packets mapping spin–orbit dynamics of the potassium $3d$ fine structure doublet. Our shaper-based approach establishes a route to fully controllable 3D light fields for chiral-sensitive light–matter interactions and ultrafast spectroscopy. 
\end{abstract}


\maketitle

\begin{figure}
	\includegraphics[width=\linewidth]{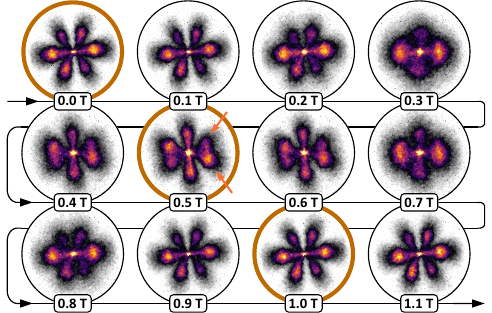}
	\caption{Time series of VMI images from the non-collinear bichromatic pump-probe study of the SOWP in the $3d$ fine structure doublet mapped into the continuum by the time-delayed blue pulse in PLP configuration. \label{fig:resultsIII}}
\end{figure}

The recognition of light as a purely transverse wave, first established by Fresnel through his studies of polarization and soon after confirmed by Maxwell’s electromagnetic theory, forms a cornerstone of modern optics. Today, light fields can be structured almost at will, enabling diverse applications across optical technologies, from super-resolution microscopy and optical tweezers to materials processing, as well as in quantum technologies for metrology, information processing and the coherent control of quantum phenomena \cite{Rice:2000,Shapiro:2003,Goswami:2003:PR:385,Tannor:2007} (see \cite{Rubinsztein-Dunlop:2016:JO:013001} for a review).\\
The generation of structured light generally involves either tailoring its spatial properties, including the generation of optical vortex beams \cite{Shen:2019:LSA:1,Forbes:2021:NP:253}, or controlling the spectro-temporal characteristics of ultrashort laser pulses \cite{Weiner:2011:OC:3669,Monmayrant:2010:JPB:103001,Koehler:2011:OE:11638}.
Ultrafast pulse shaping has advanced substantially since the advent of scalar pulse shaping \cite{Weiner:2000:RSI:1929}, which allowed temporal control of the amplitude and spectral phase of linearly polarized pulses. Subsequent developments introduced polarization shaping \cite{Brixner:2004:PRL:208301,Dudovich:2004:PRL:103003}, vector field synthesis \cite{Manzoni:2015:LPR:129} and supercontinuum pulse shaping \cite{Kerbstadt:2017:OE:12518}, permitting sub-cycle control of the vectorial electric field \cite{Ivanov:2014:NP:501}. A common feature of these techniques is that the resulting light fields are planar, with the local polarization vector confined to the transverse plane normal to the propagation direction.\\
In contrast, three-dimensional (3D) light fields, featuring an additional longitudinal electric field component, are generated, for example, by tight focusing \cite{Bauer:2015:Science:964,Ayuso:2021:Optica:1243} or the non-collinear superposition of laser beams. The latter approach is widely used in non-linear optics, including ultrashort pulse characterization \cite{Monmayrant:2010:JPB:103001},  coherent anti-Stokes Raman scattering \cite{Druet:1981:PQE:1} and non-collinear optical parametric amplification \cite{Cerullo:2011:LPR:323}. In the extreme non-linear optical regime, Hickstein \textit{et al.} demonstrated control over the polarization state of XUV pulses by driving the high harmonic generation (HHG) with two non-collinearly overlapping counter-rotating circularly polarized (CRCP) infrared pulses \cite{Hickstein:2015:NP:743}. Recently, 3D fields generated by the non-collinear superposition of two pulsed beams with variable polarization and spectrum were proposed as efficient enantio-sensitive probes \cite{Ayuso:2019:NP:866,Neufeld:2019:PRX:031002} of chiral light-matter interactions \cite{Wanie:2024:Nature:109,Sparling:2025:PCCP:2888,Ayuso:2022:PCCP:26962}. Theoretical work has shown that such light fields are locally chiral, offering enhanced sensitivity in interactions with chiral matter, such as chiral molecules. The exceptionally strong chiroptical responses predicted for these 3D fields have attracted significant interest, positioning them as promising candidates for chiral recognition applications \cite{Neufeld:2021:PRR:L032006,Habibović:2024:NRP:663}.
It has also been proposed that locally chiral light fields could imprint chirality on atomic systems by exciting unusual electronic superposition states \cite{Ordonez:2019:PRA:043416,Mayer:2022:PRL:243201} and create chiral photoelectron wave packets \cite{Katsoulis:2022:PRA:043109,Geyer:2025:PRR:L032061}.\\
Here, we report the first experimental realization of femtosecond (fs) 3D polarization-tailored bichromatic laser fields. Using these 3D fields, we create free-electron angular momentum superposition states via atomic resonance-enhanced multiphoton ionization (REMPI), yielding photoelectron wave packets that have not been observed previously. In atomic systems, access to three independent field components unlocks all dipole selection rules $\Delta m=0,\pm1$, thus opening up excitation pathways to target channels that are inaccessible to planar laser fields. Shaping of the 3D field allows us to control the quantum interference of these pathways, leading to the generation of photoelectron superposition states spanning all magnetic quantum numbers.\\
In our experiment, the 3D fields are generated by the non-collinear superposition of two polarization-shaped fs-laser pulses of different colors, provided by a supercontinuum polarization-pulse shaper \cite{Kerbstadt:2017:OE:12518,Koehnke:2023:NJP:123025}. The resulting photoelectron wave packets, created through the interaction of the 3D fields with potassium atoms are detected with energy and angular resolution using velocity-map imaging (VMI) \cite{Eppink:1997:RSI:3477}. Analysis of the measured photoelectron momentum distributions (PMDs), by comparison with simulations, confirms that they arise from a coherent superposition of partial waves encompassing all magnetic quantum numbers. Furthermore, we demonstrate the suitability of the non-collinear ionization scheme for time-resolved pump-probe studies by imaging ultrafast spin-orbit wave packet (SOWP) dynamics in the potassium $3d$ fine structure doublet. The observed PMDs highlight the potential of the 3D fields for time-resolved investigations of ultrafast dynamics with unprecedented directional sensitivity. Specifically, our shaper-based approach provides a direct route to the generation of fully controlled 3D fields and paves the way for experimental realization of the chiroptical schemes proposed by Smirnova \textit{et al.} \cite{Ayuso:2019:NP:866}, including the creation of chiral electron wave packets in atoms \cite{Mayer:2022:PRL:243201} and molecules \cite{Chen:2024:NC:565}, and the efficient enantio-sensitive discrimination of chiral molecules \cite{Ayuso:2022:PCCP:26962,Davino:2025:PRA:014022} using locally chiral light fields. \\
\begin{figure}
	\includegraphics{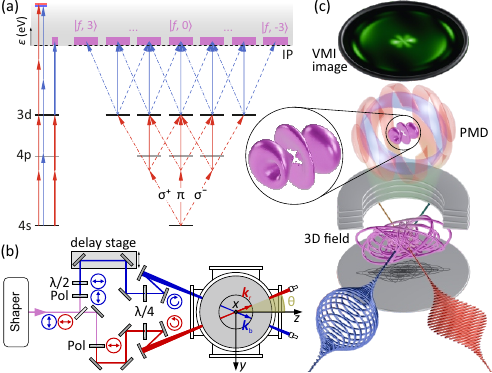}
	\caption{(a) Simplified (left) and detailed (right) bichromatic excitation scheme for the potassium atom interacting with a 3D laser field.
	(b) Schematic of the experimental setup including the pulse shaper, the non-collinear interferometer and the VMI spectrometer.
	(c) Illustration of the experiment.\label{fig:setup}}
\end{figure}
We consider the interaction of a potassium atom with a bichromatic 3D laser field. The 3D field is generated by the non-collinear superposition of two pulses of different colors, referred to as red and blue pulses. The pulses propagate in the $y$-$z$-plane along the wave vectors $\vect{k}_\mathrm{r}$ and $\vect{k}_\mathrm{b}$, oriented at $\pm\qty{22.5}{\degree}$ relative to the $z$-axis. At the intersection region, the resulting temporal 3D field is the sum of the individual fields $\vect{E}_{\vect{k}_\mathrm{r}}(t)$ and $\vect{E}_{\vect{k}_\mathrm{b}}(t)$ as
\begin{equation} \label{eq:3DField_general}
	\vect{E}(t) = \vect{E}_{\vect{k}_\mathrm{r}}(t) + \vect{E}_{\vect{k}_\mathrm{b}}(t),
\end{equation}
which oscillates locally in all three spatial dimensions.\\
The central wavelengths of the red and blue pulses are $\lambda_\mathrm{r}=\qty{929}{nm}$ and $\lambda_\mathrm{b}=\qty{720}{nm}$, respectively. The resulting field ionizes the potassium atom through multiple ionization pathways, as illustrated in Fig.~\ref{fig:setup}(a). Individually, the red and blue pulses induce four- and three-photon ionization. The corresponding single-color photoelectron wave packets are centered at $\varepsilon_\mathrm{r}=\qty{1.0}{eV}$ and $\varepsilon_\mathrm{b}=\qty{0.8}{eV}$, defining a broad energy window for background-free detection of frequency mixing signals. Photoelectron wave packets created by the interplay of the two pulses, are released along different three-photon frequency mixing pathways. Our choice of central wavelengths ensures that the pathway involving two red and one blue photon is resonantly enhanced by the potassium $3d$-state and thus dominates the photoelectron spectrum. The corresponding photoelectron wave packet from (2+1) REMPI is centered at $\varepsilon_{\text{3D}}=\qty{0.05}{eV}$, well-separated from the two single-color signals.\\
In the spherical basis, the 3D field is described by two CRCP transverse components in the $x$-$y$-plane and one linearly polarized longitudinal component along the $z$-axis. Within the dipole approximation, the transverse components drive $\sigma^{\pm}$ transitions, while the longitudinal component drives $\pi$ transitions. 
Figure~\ref{fig:setup}(a) (right) illustrates the manifold of ionization pathways for the (2+1) REMPI process, including all dipole-allowed transitions between the magnetic sublevels governed by the selection rules $\Delta m = 0,\pm1$.  
Unlike planar fields, the 3D configuration provides access to all seven angular-momentum continua $|f,m\rangle$ ($m=-3,...,+3$). According to Fano's propensity rule \cite{Fano:1985:PRA:617}, the total photoelectron wave function for $N$-photon ionization from the $l=0$ ground state is expressed as a superposition of angular momentum partial waves
\begin{equation} \label{eq:ContiWF_general}
	\psi^{(N)}(\varepsilon,\vect{\Omega}) =
	\sum_{m=-N}^N\limits
	a_{N,m}\left(\varepsilon\right)
	\operatorname{Y}_{N,m}(\vect{\Omega}).
\end{equation}
Here, the amplitudes $a_{N,m}\left(\varepsilon\right)$ define the kinetic energy spectrum of the partial waves and their angular part is given by the spherical harmonics $\operatorname{Y}_{N,m}(\vect{\Omega})$.\\
In our experiment, \qty{20}{\fs}, \qty{790}{\nm} pulses with energies of \qty{0.9}{\mJ} from an amplified fs-laser system are coupled into a neon-filled hollow-core fiber to generate a white-light supercontinuum. The white-light pulses are compressed and converted into orthogonally linearly polarized (OLP) bichromatic pulses using a home-built pulse shaper \cite{Kerbstadt:2017:OE:12518,Kerbstadt:2019:NC:658}. A non-collinear interferometer, shown in Fig.~\ref{fig:setup}(b), separates the red and the blue pulse via orthogonal polarizers. Additional $\lambda/2$- and $\lambda/4$-waveplates in each arm enable the generation of bichromatic pulses with parallel linear (PLP), corotating circular (COCP), CRCP, red-circular/blue-linear (RCBL) or red-linear/blue-circular (RLBC) polarization. Two spherical mirrors ($f=\qty{250}{\mm}$) focus the beams into the interaction region of a VMI spectrometer, where they intersect at an angle of $2\theta = 45^{\circ}$. The photoelectron imaging setup is shown in Fig.~\ref{fig:setup}(c). An einzel lens in the VMI setup enables operation in momentum- or spatial-imaging mode. Further details of the setup are provided in \cite{Kerbstadt:2019:NC:658}.\\
\begin{figure}
	\includegraphics{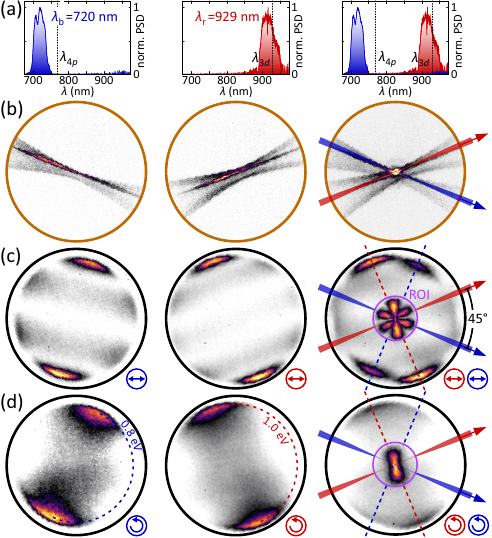}
	\caption{(a) Measured optical spectra.
		(b) VMI images recorded in spatial imaging mode for the colors individually and combined.
		VMI images recorded in momentum imaging mode for the (c) PLP and (d) COCP configuration for blue pulse (left), red pulse (center) and both pulses combined (right). The two-color signal created by the 3D fields is indicated by purple circles marking the ROI. \label{fig:characterization}}
\end{figure}
Figure~\ref{fig:characterization} shows representative results, illustrating the experimental setup. Panel (a) shows measured power spectral densities (PSDs) of the red and blue pulses. Panel (b) displays spatial images of the individual beams and their intersection. The (2+1) REMPI signal appears as a bright spot at the beam intersection and is used for \textit{in-situ} optimization of the spatio-temporal overlap. Panels (c) and (d) present the corresponding PMDs for the PLP and COCP polarization configurations, respectively. The observed $f$- and $g$-type wave packets created by the individual blue or red pulses agree with the expected three- and four-photon ionization pathways. Their (rotational) symmetry axes are aligned perpendicular (PLP) and parallel (COCP) to the respective propagation directions. By design, the single-color contributions are well-separated from the two-color signal near the ionization threshold, enabling background-free detection of photoelectron wave packets created by the frequency mixing of both pulses. In the following, we focus on this region of interest (ROI), marked by the purple circle, and analyze the PMDs created by different 3D field configurations.\\
\begin{figure*}[t]
	\includegraphics[width=\linewidth]{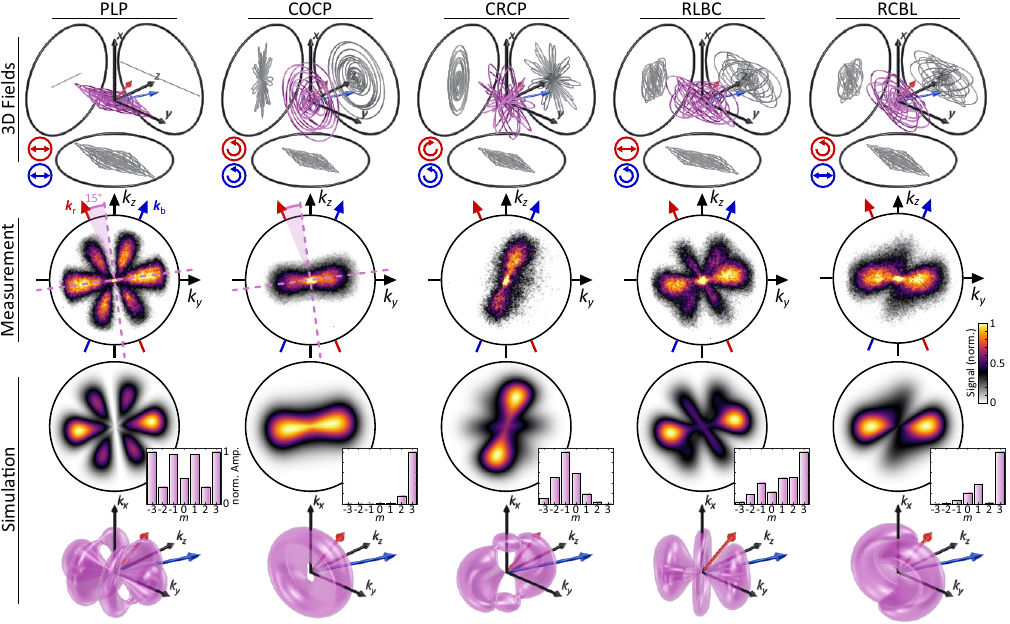}
	\caption{
		First row: 3D fields for the PLP, COCP, CRCP, RLBC and RCBL configuration.
		Second row: Measured VMI images within the ROI. 
		Third and forth row: Simulated 2D projections and 3D densities of the PMDs. The insets show the decomposition of the wave packets into angular momentum partial waves. \label{fig:resultsI}}
\end{figure*}
Our key results are presented in Fig.~\ref{fig:resultsI}.
The top row shows the 3D laser fields (purple) along with their cartesian projections (grey) with red and blue arrows indicating the corresponding propagation directions. 
The second row shows the measured VMI images within the ROI, which are compared to simulated photoelectron projections (third row) and the corresponding 3D PMDs (fourth row) obtained from Eq.~\eqref{eq:ContiWF_general}. The close agreement across all five configurations validates our model and provides the basis for a more detailed analysis. The bar plot insets next to the 3D PMDs display the $m$-partial wave amplitudes (cf. Eq.~\eqref{eq:ContiWF_general}) in the common reference frame defined by the $z$-axis oriented along the bisector of the two propagation axes. The $m$-decompositions demonstrate that all target states are accessed via $\sigma^\pm$ and $\pi$ transitions.
In the following, we examine the five polarization configurations column by column.\\
In the PLP configuration, the field oscillates within the $y$-$z$-plane and is therefore equivalent to a collinear bichromatic field with tilted linear polarizations propagating along the $x$-direction. Its projections along the $x$-axis form a Lissajous-type curve. The PMD resembles a deformed $|f,0\rangle$-type wave, packet rotated by approximately \SI{15}{\degree} towards the red propagation axis.\\
The PMD for the COCP configuration shows a predominantly $|f,3\rangle$-type wave packet and is aligned in the same direction. In addition to the prominent $m=3$ component, it contains smaller adjacent $m$ contributions. The toroidal photoelectron emission mirrors the structure of the 3D field: the field vector traces a circular trajectory around $z$, while gradually tilting around $y$, giving rise to the characteristic petaloid ($y$) and helical ($z$) projections.\\
Switching the handedness of the red pulse yields the CRCP pulse configuration (third column) where the characteristic projection exchange roles: rather than rotating around the $z$-axis, the field now circles predominantly around the $y$-axis while tilting around the $z$-axis. As a result, the photoelectron emission is directed mainly along $z$ in the CRCP configuration. The $m$-decomposition reveals a broader distribution of partial waves, with its maximum centered around $m=-1$.\\
For the discussion of the RLBC and RCBL configurations, we adopt a different reference frame aligned with the red pulse propagation direction. 
In our 2+1 REMPI scheme, a single ionization pathway is strongly enhanced, allowing us to choose a suitable reference frame such that a single $|3d,m\rangle$-state is excited by the red pulse, either by two $\sigma^\pm$- or two $\pi$-transitions. The non-collinear ionizing blue pulse drives one-photon ionization from the excited state to three different target continua $|f,m\rangle$ with $\Delta m = 0,\pm1$. The bound-state excitations in the RLBC and RCBL configurations are identical to the PLP and COCP case, respectively: in the reference frame defined by the red pulse, the two-photon excitation populates selectively either the $|3d,0\rangle$ (RLBC) or the $|3d,2\rangle$ (RCBL) state. The pronounced differences in the PMD structures originate from the ionization step, where the non-collinear blue pulse drives transitions into a set of adjacent $m$-continua with $m=-1,0,1$ (RLBC) and $m=1,2,3$ (RCBL). \\
The previously unobserved PMD structures provide a glimpse into the capabilities of advanced coherent control unlocked by full vectorial 3D fields, which enables simultaneous access to all  $\Delta m$ selection rules. The resulting $m$-distributions in the common reference frame reflect this capability through their richer and more intricate structure. Our results show that full experimental access to these unusually structured 3D PMDs will require the use of emerging 3D reconstruction approaches that combine VMI with high-resolution time-of-flight detection \cite{Davino:2023:RSI:013303}.\\
To demonstrate the capability of our 3D non-collinear scheme for imaging of ultrafast electron dynamics in a pump–probe setting, we applied it to map a SOWP from the $\left\{3d_{3/5},3d_{5/2}\right\}$ fine structure doublet into the continuum. In our (2+1) REMPI scheme, the red pulse launches the SOWP, while the delayed blue pulse probes its evolution. By varying the delay in increments of $0.1~T$ with $T=\qty{14.45}{\pico\second}$, we recorded ten PMDs per SOWP period.\\
Figure~\ref{fig:resultsIII} shows the PLP results. At the beginning of each period ($0.0~T$), the PMD corresponds to the rotated $|f,0\rangle$-type wave packet discussed above (cf. Fig.~\ref{fig:resultsI}). 
As the delay increases, the photoelectron distribution changes, mapping the precession of the coupled electron spin and orbital alignment about the total angular momentum \cite{Zamith:2000:EPJD:255,Bayer:2019:NJP:033001}.
After half a period ($0.5~T$), the PMD has evolved into the the characteristic shape of a rotated $|f,\pm1\rangle$-type photoelectron wave packet, indicating the transition $|3d,0\rangle\rightarrow|3d,\pm1\rangle$ in the bound-system. 
This change in the angular part of the wave function is a hallmark for SOWP dynamics and reflects the accompanying flip of the electron spin induced by the spin–orbit interaction \cite{Bayer:2019:NJP:033001}.
The subsequent return to the original $|f,0\rangle$-type distribution at $1.0~T$ completes the spin-orbit cycle.\\
Notably, the PMDs exhibit a pronounced asymmetry in their lobes (orange arrows), reflecting directional sensitivity enabled by the non-collinear scheme. 
By incorporating dynamics as an additional degree of freedom, our 3D non-collinear scheme enables the creation of more intricate PMDs and establish the potential of 3D vectorial laser field control for time-resolved studies.\\
In conclusion, we have experimentally demonstrated a versatile non-collinear bichromatic framework that enables coherent control of MPI through full vectorial control of 3D light fields. By non-collinearly superimposing two polarization-tailored pulses, we created 3D fields with both transverse and longitudinal components, which unlock all dipole selection rules in the (2+1) REMPI scheme. Our results reveal previously unobserved PMDs, which arise from the coherent superposition of photoelectron angular momentum partial waves encompassing all magnetic quantum numbers -- MPI channels that are fundamentally inaccessible to planar laser fields. The conceptual clarity of our experimental approach, combined with reference measurements using single-color ionization and fully differential energy- and angle-resolved photoelectron detection, allowed us to unambiguously identify the contributing ionization pathways. The excellent agreement between experimental data and simulations provides evidence for coherent control of MPI using 3D light fields. We further extended the approach to a non-collinear pump-probe experiment, mapping the evolution of an ultrafast SOWP in the potassium $3d$ fine structure doublet, highlighting the potential of the 3D fields for time-resolved investigations of ultrafast dynamics with unprecedented directional sensitivity. While 3D light fields have been the subject of extensive theoretical investigations, their experimental realization has been lacking. This work provides the first experimental demonstration of MPI driven by fully controllable 3D light fields. Our scheme opens new avenues for chiral recognition, advanced coherent control, and symmetry-breaking excitation schemes facilitating the creation of chiral electronic bound and continuum states.

\begin{acknowledgments}
We gratefully acknowledge financial support from the Deutsche Forschungsgemeinschaft (project no. 467215508), as well as the collaborative research program “Dynamics on the Nanoscale” (DyNano) and the Wissenschaftsraum “Elektronen-Licht-Kontrolle” (elLiKo) funded by the Niedersächsische Ministerium für Wissenschaft und Kultur.
\end{acknowledgments}
\bibliography{ultra_db.bib}
\end{document}